# First-principles study of electronic structure, optical and phonon properties of α-ZrW$_2$O$_8$

Jinping Li, [1, 3, †] Songhe Meng, [1] Cheng Yang, [1] and Hantao Lu, [2, 3]

([1] Center for Composite Materials and Structure, Harbin Institute of Technology, Harbin 150080, China)

([2] Center for Interdisciplinary Studies & Key Laboratory for Magnetism and Magnetic Materials of the MoE, Lanzhou University, Lanzhou 730000, China)

([3] Yukawa Institute for Theoretical Physics, Kyoto University, Kyoto 606-8502, Japan)

[†] Corresponding author: lijinping@hit.edu.cn

**Abstract**: ZrW$_2$O$_8$ exhibits isotropic negative thermal expansions over its entire temperature range of stability, yet so far its physical properties and mechanism have not been fully addressed. In this article, the electronic structure, elastic, thermal, optical and phonon properties of α-ZrW$_2$O$_8$ are systematically investigated from first principles. The agreements between the generalized gradient approximation (GGA) calculation and experiments are found to be quite satisfactory. The calculation results can be useful in relevant material designs, e.g., when ZrW$_2$O$_8$ is employed to adjust the thermal expansion coefficient of ceramic matrix composites.

## Ⅰ. Introduction

The zirconium tungstate (ZrW$_2$O$_8$) exhibits isotropic negative thermal expansion (relatively large, $-9 \times 10^{-6}$ K$^{-1}$) over its entire temperature range of stability (from close to absolute zero up to the decomposition temperature around 1500 K). The isotropy of the expansion is back by the fact that the cubic structure of ZrW$_2$O$_8$ remains at these temperatures. [1,2] This feature makes ZrW$_2$O$_8$ not only an important example to study this type of unusual lattice dynamics, but also potentially well suited for applications in composite materials in order to reduce the composites' overall thermal expansion to near zero. [3-7]

In experiments, the phase transition, specific heat, thermal expansion of ZrW$_2$O$_8$ have been studied by some groups [8-11], especially for the temperature- and pressure-induced phase transitions, and the temperature-dependence of the thermal. [12-14] For example, it has long been noticed that α-ZrW$_2$O$_8$ has a negative coefficient of thermal expansion, α= $-9.07 \times 10^{-6}$ K within the temperature range 2-350 K [8], and it undergoes a phase transition from *P213* to *Pa3* at 448 K [8] or 428K [12], which is associated with the onset of considerable oxygen mobility. It has also been shown that under hydrostatic pressure at room temperatures, a structural phase transition, i.e., from the cubic to an orthorhombic phase, can be produced, which begins at 2.1 kbar and completes at 3.1 kbar, with a 5% volume reduction. [13] A further increase in pressure beyond 1.5 GPa can induce an amorphous phase. [14]

In theory, previous studies mainly concern on the physical mechanism of its phase transitions and negative thermal expansion. [15-18] For example, with the structure optimization at different pressures, the elastic constants of α-$ZrW_2O_8$ have been calculated by the B3LYP. It shows that the tetrahedra around tungsten atoms is much stiffer than the $ZrO_6$ octahedra, although the Zr-O bonds are quite stiffer than the W-O bonds. [15, 18] The obtained elastic constants in the athermal limit are in excellent agreement with recent experimental results obtained near 0 K. [18] The band structures of both α and γ phases of $ZrW_2O_8$ have been studied by the density-functional theory, which shows that α-$ZrW_2O_8$ has an indirect band gap of 2.84 eV, while γ-$ZrW_2O_8$ has a direct band gap of 2.17 eV. A larger total crystal bond order is found in γ-$ZrW_2O_8$. The numerical result of the bulk modulus (104.1 GPa) for the α-phase is in good agreement with experiments. [18]

However, many aspects of physical properties of $ZrW_2O_8$, in our opinion, are still unclear and have not been received a systematic investigation so far. In this paper, by employing first principles method, we perform numerical calculations on various properties of the material, including the electronic structure, mechanical, optical, thermal and phonon properties. Comparisons with experimental data are also made.

## Ⅱ. Computational methodology

The first-principles calculations are performed with plane-wave ultrasoft pseudopotential by means of GGA with Perdew-Burke-Ernzerhof (PBE) functional as implemented in the CASTEP code (Cambridge Sequential Total Energy Package). [19] The ionic cores are represented by ultrasoft pseudopotentials for Zr, W and O atoms. For Zr atom, the configuration is [Kr]$4d^25s^2$, where the $4s^2$, $4p^6$, $4d^2$ and $5s^2$ electrons are explicitly treated as valence electrons. For W atom ([Xe]$5d^46s^2$), the $5s^2$, $5p^6$, $5d^4$ and $6s^2$ electrons are treated as valence electrons. And for O atom ([He]$2s^22p^4$), $2s^2$ and $2p^4$ are valence electrons. The plane-wave cut-off energy is 380eV, and the Brillouin-zone integration is performed over the 6×6×6 grid sizes using the Monkorst-Pack method for structure optimization. This set of parameters assures the total energy convergence with the accuracy as $5.0 \times 10^{-6}$eV/atom, the maximum force 0.01eV/Å, the maximum stress 0.02 GPa, and the maximum displacement $5.0 \times 10^{-4}$Å.

In the following sections, after performing optimization on the geometric structure of α-ZrW$_2$O$_8$ by GGA, we calculate the electronic structure, elastic modulus, thermal, optical and phonon properties of α-ZrW$_2$O$_8$ systematically. The results are summarized and compared with available experimental data.

## Ⅲ. Results and discussions
### (1) Electronic Structures of α-ZrW$_2$O$_8$

The space group of α-ZrW$_2$O$_8$ is P21-3 and the local symmetry is O5h. Moreover, because of its cubic lattice structure, α-ZrW$_2$O$_8$ is fully characterized by a single lattice constant *a*. In order to check the applicability and accuracy of the ultrasoft pseudopotential, the GGA calculation of the perfect bulk α-ZrW$_2$O$_8$ is carried out to determine the optimized lattice constant *a*. The result is 0.92867nm, in good agreement with other theoretical predictions, e.g., 0.93565nm or 0.92691 by B3LYP [17], 0.91494nm by LDA [18]. It is also consistent with the experimental value, 0.9160nm. [2]

The following electronic band structure calculation by GGA (not shown here) indicates the existence of an indirect band gap, whose value $E_g$ is about 3.31eV, larger than other theoretical result, e.g., 2.84eV by LDA,[18] more coincides with the experimental value, i.e., 4.0eV, [20], which is smaller than the experimental value due to the well-known underestimate of conduction-band energies in ab initio calculations.

The total density of states (DOS) is presented in Fig. 1(a). There are two parts in the valence bands, *i.e.*, the lower region from -17.5eV to -14.9eV, and the upper region from -5.3eV to 0.7eV. As shown in Fig. 1(a), the lower part of the valence bands is predominated by O 2*s*; while the upper part largely consist of O 2*p*, accompanied with a small amount of hybridization with Zr 4*d* and W 5*d*, as shown in Figs. 1(b)-(d). The conduction bands below 10eV are mostly composed of Zr 4*d* and W 5*d* with some amount of O 2*p*. In higher energy regime, the contributions from the 5*s*, 4*p* orbitals of Zr atom, and the 6*s*, 5*p* orbitals of W atom, are dominant.

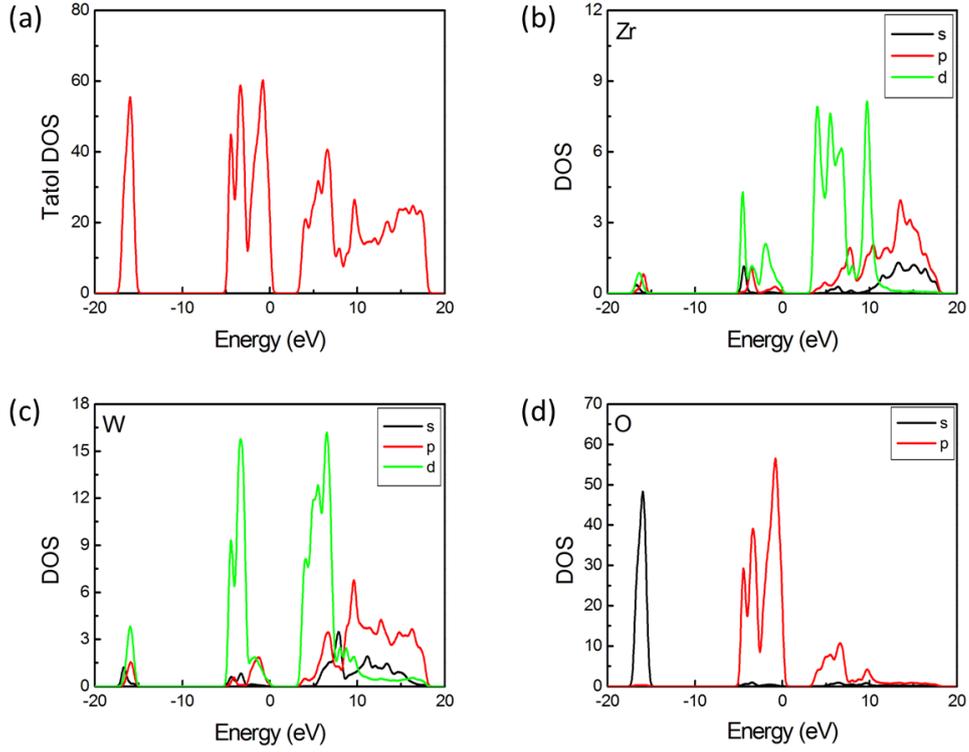

Fig.1. The total density of state (DOS) and partial DOS of Zr, W, O atom are shown in (a), (b), (c) and (d), respectively.

### (2) Elastic properties of α-ZrW$_2$O$_8$

We performed calculations on mechanical properties of α-ZrW$_2$O$_8$ by GGA-PBE within the ultrasoft pseudopotential. After cubic structure geometry optimization, the final cell parameter *a* is 0.92867nm, well agreeing with experimental value, 0.9160nm. [2] Table I shows the results on mechanical parameters of α-ZrW$_2$O$_8$, including elastic modulus ($C_{11}$, $C_{12}$ and $C_{44}$), volume modulus (*B*), shear modulus (*G*), Young's modulus (*E*), Poisson's ratio (*v*) and Debye temperature ($\Theta_D$). A comparison with experiments is also made in the table. We can see that the numeric results coincide well with the experimental data. [21]

Table I. Mechanical properties of α-ZrW$_2$O$_8$ obtained by GGA (the first line).

A comparison with experimental data is made in the second line.

| $C_{11}$(GPa) | $C_{12}$(GPa) | $C_{44}$(GPa) | B(GPa) | E(GPa) | G(GPa) | v | $\Theta_D$(K) |
|---|---|---|---|---|---|---|---|
| 141.23 | 67.82 | 28.11 | 92.29 | 97.23 | 31.55 | 0.324 | 314.7 |
| 161.8 | 75.5 | 29.4 | 104.3 | 98.8 | 36.8 | 0.342 | 333 |

The theoretical result of Debye temperature, as shown in the last column of Table I, can be obtained in the following. First note that the compress wave velocity $v_i$ and shear wave velocity $v_s$ can be calculated from the modulus:

$$v_i = \sqrt{\frac{B_H + \frac{4}{3}G_H}{\rho}} \quad ; \quad v_s = \sqrt{\frac{G_H}{\rho}} \qquad (1)$$

where $B_H$ is volume module, $G_H$ is shear module, and $\rho$ is the theoretical density. The average speed of sound $v_m$ can be defined as

$$v_m = \left[\frac{1}{3}\left(\frac{2}{v_s^3} + \frac{1}{v_i^3}\right)\right]^{-1/3} \qquad (2)$$

Then the Debye temperature can be obtained by the formula:

$$\Theta_D = \frac{h}{k_B}\left[\frac{3n}{4\pi}\left(\frac{N_A \rho}{M}\right)\right]^{1/3} v_m \qquad (3)$$

where $h$ is Planck constant, $k_B$ is Boltzmann constant, $N_A$ is Avogadro's constant, $n$ is atom number of per primitive cell, and $M$ is molecule mass of per primitive cell. By using the above equations, we can determine the Debye temperature as 314.7K, which is well consistent with experimental value [21] (the deviation is about 5.50%).

### (3) Thermal conductivity of α-ZrW$_2$O$_8$

Zirconium tungstate, ZrW$_2$O$_8$, has received considerable attention in recent years because of its isotropic negative thermal expansion over a wide temperature range from 4 to 1050 K. It is well known that Young's modulus has a close relationship with thermal conductivity, as higher Young's modulus can help to increase the thermal conductivity. The theoretical minimum thermal conductivity and its underlying origin have been discussed, and there are two models which can produce the lower limit of the thermal conductivity of a crystal. One is based on the Debye model. Clarke [22] suggests that the minimum can be obtained after replacing different atoms by an equivalent atom with a mean atomic mass $M/m$:

$$k_{min} = 0.87 k_B \left(\frac{M}{m \cdot N_A}\right)^{-2/3} E^{1/2} \rho^{1/6} \qquad (4)$$

Another is the Cahill's model [23], which relates the minimum thermal conductivity to materials

sound velocity and the density of number of atoms per volume, as shown in Eq. (5):

$$k_{\min} = \frac{k_B}{2.48} \rho^{2/3} (v_i + 2v_s) \quad (5)$$

In the above two equations, $E$ is the elastic modulus (Young's modulus), $\rho$ is the density, $M$ is the molar mass, $m$ is the total number of atoms per formula, $k_B$ is Boltzmann's constant, $N_A$ is Avogadro's number, and $p$ is the density of number of atoms per volume. These models have been validated and widely used in searching for materials as thermal barrier coatings. [22]

In the case of α-ZrW$_2$O$_8$, the resulting thermal conductivity is 0.806 W m$^{-1}$ K$^{-1}$ from Cahill's model, and 0.781 W m$^{-1}$ K$^{-1}$ from the Clark's model, well agreeing with the experimental value, i.e., approximately 0.80 W m$^{-1}$ K$^{-1}$.[24] Generally, the quantity obtained by Cahill's model is larger than that by Clark's model.

## (4) Optical properties of α-ZrW$_2$O$_8$

The complex dielectric function ε contains a real part $\varepsilon_1$ and an imaginary part $\varepsilon_2$. The imaginary part can be calculated from the band structure directly by taking into account of interband transitions, while the real part can be obtained by the following use of the Kramers-Kronig relations. [25] Other optical properties can be computed from the complex dielectric function. [25] Figures 2(a) and (b) show the complex dielectric function and refractive index as a function of photon energy, obtained by GGA. In Fig. 2(a), we can see that the static dielectric constant ($\varepsilon_1(E=0)$) is 3.93eV, which is actually quite close to the experimental value of cubic ZrO$_2$, 4.1eV.[26] The static refractive coefficient ($n_0$) obtained by GGA is about 1.98. And for cubic ZrO$_2$, the corresponding value in experiments is 2.07 [27] or 1.932 [28]. According to the above, we can speculate that the optical properties of ZrW$_2$O$_8$ are much similar to those of cubic ZrO$_2$.

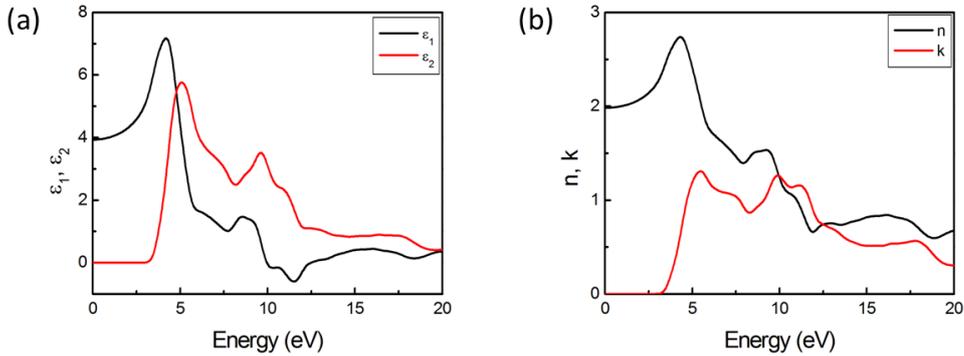

Fig.2. (a) The dielectric function and (b) the refractive index of α-ZrW$_2$O$_8$ obtained by GGA.

## (5) Phonon properties of α-ZrW$_2$O$_8$

The low-energy part of the phonon dispersion (a) and its DOS (b) of α-ZrW$_2$O$_8$ obtained by GGA are presented in Fig. 3. The Bradley-Cracknell notation is used for the high-symmetry points along which the dispersion relations are obtained, e.g., G=(0, 0, 0), M=(0.5, 0.5, 0) and R=(0.5, 0.5, 0.5). In the cell of ZrW$_2$O$_8$, there are 44 atoms including 4 Zr atoms, 8 W atoms and 32 O atoms, which produce 132 vibration modules, including 3 acoustic branches and 129 optical branches. The acoustic branches exist only in the low-frequency regime, and approach zero frequency towards the G point, as shown in Fig. 3(a). No imaginary phonon frequencies are found in ZrW$_2$O$_8$, which indicates that it can be stable at room temperature and normal pressure. The low energy range of phonon dispersion contains large number of non-dispersive phonon branches, which give rise to several peaks in density of states. The lowest optical mode is calculated at 1.243THz (namely, 41.5cm$^{-1}$) , whose error is about 3.75% compared with the experimental value of 40cm$^{-1}$ from Raman [2] as well as infra-red measurements. [10]

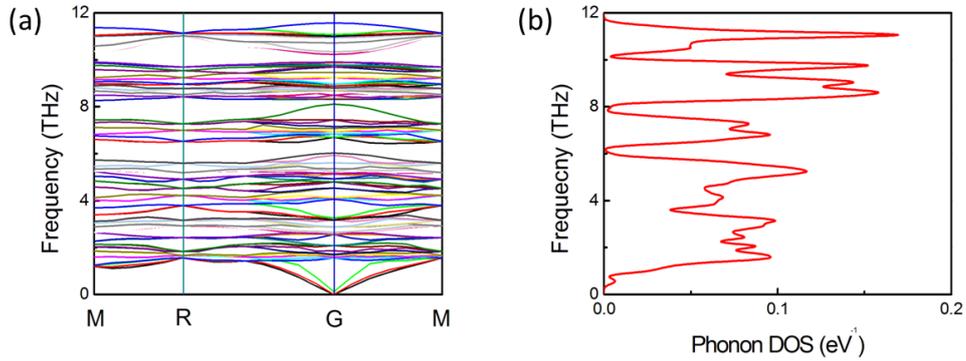

Fig.3. Phonon dispersion (a) and phonon DOS (b) of α-ZrW$_2$O$_8$ obtained by GGA

When the acoustic vibration is concerned, it has been shown that the frequency can be related to the bond stiffness and atom mass in the following way:

$$\omega = \sqrt{\frac{K}{m_{red}}} = \sqrt{K(\frac{1}{m_1} + \frac{1}{m_2})} \qquad (6)$$

where $\omega$ is the acoustic vibration frequency, $K$ the bond stiffness, $m_{red}$ is the equivalent atom mass, $m_1$ and $m_2$ are the mass of two atoms within one bond, respectively.

From the above formula, we can estimate the contributions of various bonds to the acoustic vibrations. For example, since the stiffness of O-Zr bonds is bigger than that of W-O bonds, [15, 18] and the oxygen mass is the smallest among the three elements, it is expected that the optical vibration module of the O atoms mainly contribute to the high-frequency part. On the other hand, the stiffness of W-O bonds is weaker and the mass of the W atoms is the biggest, thus the optical vibration frequency of the W atoms should be the lowest. As for Zr atoms, the corresponding frequencies are located between those of W and O atoms. Besides, as shown in Fig. 3, there is no phonon band gap because the three acoustic branches merge into the low-lying part of optical branches.

The thermal dynamics and the heat capacity of the $ZrW_2O_8$ are also calculated by GGA, which are presented in Fig. 4. Figure 4(a) shows the temperature dependence of enthalpy, free energy and entropy (multiplied by temperature) over a range of 0-1000K. The enthalpy and entropy increase with temperature. Since the ions polarize each other when they close each other, the relationship line between entropy and temperature presents certain bending. Figure 4(b) shows the temperature dependence of heat capacity from the phonon contributions. In high temperatures (up to 1000K), the heat capacity approaches the so-called Dulong-Petit limit, i.e., $C=3Nk_B$.

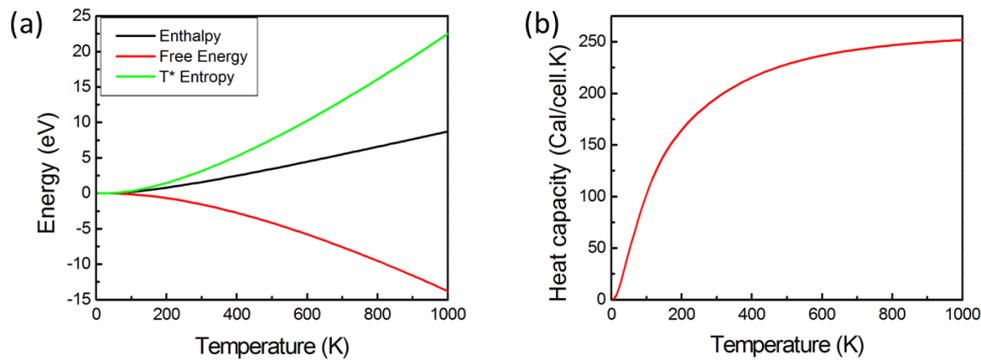

Fig.4. Thermodynamics (a) and heat capacity (b) of α-$ZrW_2O_8$ obtained by GGA.

Figure 5 (a) shows the results of the Debye temperature of α-$ZrW_2O_8$. It can be seen that the Debye temperature first decreases at low temperatures, reaching a small dip, and then increases monotonically to approach a maximum (903.04K at 1000K). The dip locates at T=15K, with corresponding Debye temperature as 247.17K, whose origin is still under discussions. At T=5K, the Debye temperature is 317K, well agreed with the experimental results, 333K [21] or 311K [29]. Figure 5 (b) shows the comparison of the specific heat divided by temperature between the theoretical calculations and experimental data. [29] The numerical results are well agreed with the experiments.

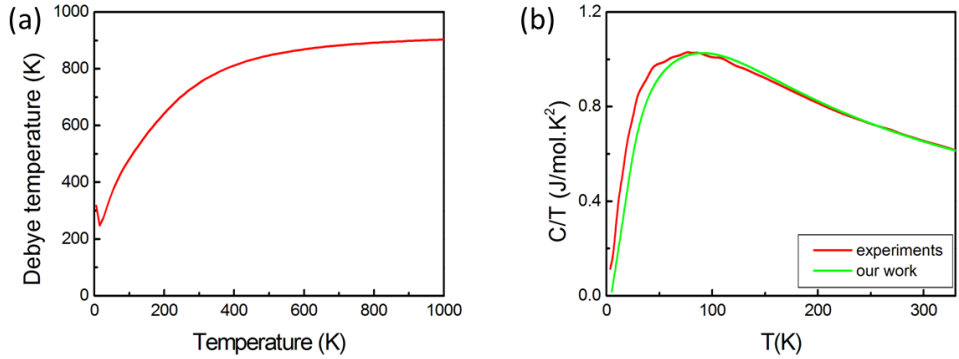

Fig.5. (a) Debye Temperature of α-$ZrW_2O_8$ obtained by GGA; (b) the comparison of the specific heat divided by temperature between the theory and experiments.

## Ⅳ. Conclusion

The electronic structure, elastic, optical and phonon properties of α-$ZrW_2O_8$ have been systematically studied by first-principles generalized gradient approximation (GGA). The results can be summarized as follows. The electronic band structure calculation shows that $ZrW_2O_8$ is a semiconductor material with an indirect band-gap, whose band gap is 3.305 eV. The elastic modules, Poisson's ratio and Debye temperature obtained by GGA are well agreed with experimental data. The minimum thermal conductivity, depending on which model we used, either the Cahill's model or Clark's model, is 0.806 W·$m^{-1}$·$K^{-1}$ or 0.781 W·$m^{-1}$·$K^{-1}$. For the optical properties, the static dielectric constant ($\varepsilon_1(E=0)$) and the static refractive coefficient ($n_0$) are 3.93eV and 1.98, respectively, very close to those of cubic $ZrO_2$. The phonon properties are investigated in detail, including its dispersions and some related thermodynamic quantities. The temperature dependence of the heat capacity divided by temperature coincides well with the experimental results.

## References

1. C. A. Perottoni and J. A. H. da Jornada, "Pressure-induced amorphization and negative thermal expansion in $ZrW_2O_8$," *Science*, **280** [5365] 886-889 (1998).

2. T. A. Mary, J. S. O. Evans, T. Vogt, and A. W. Sleight, "Negative thermal expansion from 0.3 to 1050 Kelvin in $ZrW_2O_8$," *Science*, **272** [5258] 90-92 (1996).

3. Y. Yamamura, N. Nakajima, and T. Tsuji, "Calorimetric and x-ray diffraction studies of α-to-β structural phase transitions in HfW$_2$O$_8$ and ZrW$_2$O$_8$," *Phys. Rev. B*, **64** [18] 184109 (2001).

4. D. A. Fleming, D. W. Johnson, and P. J. Lemaire, "Article comprising a temperature compensated optical fiber refractive index grating"; *U.S. Patent* 5,694,503, 1997.

5. H. Liu, Z. Zhang, X. Cheng, and J. Yang, "Thermal expansion of ZrO$_2$- ZrW$_2$O$_8$ composites prepared using co-precipitation route," *Int. J. Mod Phys B*, **23** [06n07] 1449-1454 (2009).

6. L. Sun and P. Kwon, "ZrW$_2$O$_8$/ZrO$_2$ composites by in situ synthesis of ZrO$_2$+WO$_3$: processing, coefficient of thermal expansion, and theoretical model prediction," *Mater. Sci. Eng., A*, **527** [1-2] 93-97 (2009).

7. L. Sun, A. Sneller, and P. Kwon, "ZrW$_2$O$_8$-containing composites with near-zero coefficient of thermal expansion fabricated by various methods: comparison and optimization," *Compos. Sci. Technol.*, **68** [15] 3425-3430 (2008).

8. J. S. Evans, W. I. F. David, and A. W. Sleight, "Structural investigation of the negative-thermal-expansion material ZrW$_2$O$_8$," *Acta Cryst. B*, **55** [3] 333-340 (1999).

9. G. Ernst, C. Broholm, G. R. Kowach, and A. P. Ramirez, "Phonon density of states and negative thermal expansion in ZrW$_2$O$_8$," *Nature*, **396** [6707] 147-149 (1998).

10. A. P. Ramirez and G. R. Kowach, "Large low temperature specific heat in the negative thermal expansion compound ZrW$_2$O$_8$," *Phys. Rev. Lett.*, **80** [22] 4903 (1998).

11. Y. Yamamura, N. Nakajima, and T. Tsuji, "Heat capacity anomaly due to the a-to-b structural phase transition in ZrW$_2$O$_8$," *Solid State Communications*, **114** [119] 453-455 (2000).

12. J. S. O. Evans, T. A. Mary, T. Vogt, M. A. Subramanian, and A. W. Sleight, "Negative thermal expansion in ZrW$_2$O$_8$ and HfW$_2$O$_8$," *Chem. Mater.*, **8** [12] 2809-2823 (1996).

13. Z. Hu, J.D. Jorgensen, S. Teslic, S. Short, D.N. Argyriou, J.S.O. Evans, and A.W. Sleight, "Pressure-induced phase transformation in ZrW$_2$O$_8$ -compressibility and thermal expansion of the orthorhombic phase," *Physica B*, **241** 370-372 (1997).

14. L. Ouyang, Y. N. Xu, and W. Y. Ching, "Electronic structure of cubic and orthorhombic phases of ZrW$_2$O$_8$," *Phys. Rev. B*, **65** [11] 113110 (2002).

15. D. Cao, F. Bridges, G. R. Kowach, and A. P. Ramirez, "Frustrated soft modes and negative thermal expansion in ZrW$_2$O$_8$," *Phys. Rev. Lett.*, **89** [21] 215902 (2002).

16. J. S. Evans, Z. Hu, J. D. Jorgensen, D. N. Argyriou, S. Short, and A. W. Sleight, "Compressibility, phase transitions, and oxygen migration in zirconium tungstate, $ZrW_2O_8$," *Science*, **275** [5296] 61-65 (1997).

17. X. Yang, X. Cheng, X. Yan, J. Yang, T. Fu, and J. Qiu, "Synthesis of $ZrO_2/ZrW_2O_8$ composites with low thermal expansion," *Compos. Sci. Technol.*, **67** [6] 1167-1171 (2007).

18. C. A. Figueirêdo and C. A. Perottoni, "B3LYP density functional calculations on the ground-state structure, elastic properties, and compression mechanism of α-$ZrW_2O_8$," *Phys. Rev. B*, **75** [18] 184110 (2007).

19. M. D. Segall, P. J. D. Lindan, M. J. Probert, C. J. Pickard, P. J. Hasnip, S. J. Clark, and M. C. Payne, "First-principles simulation: ideas, illustrations and the CASTEP code," *J. Phys.: Condens. Matter*, **14** [11] 2717 (2002).

20. L. Jiang, H. Liu, J. Yuan, W. Shangguan, "Hydrothermal preparation and photocatalytic water splitting properties of $ZrW_2O_8$," *Journal of Wuhan University of Technology-Mater. Sci.* **25**, 919-923 (2010).

21. R. Drymiotis, H. Ledbetter, J. B. Betts, T. Kimura, J.C. Lashley, A. Migliori, A. P. Ramirez, G. R. Kowach, J. Van Duijn, "Monocrystal elastic constants of the negative-thermal-expansion compound zirconium tungstate ($ZrW_2O_8$)", *Phys Rev Lett.,* **93**, 025502 (2004).

22. D. R. Clarke, "Materials selection guidelines for low thermal conductivity thermal barrier coatings," *Surface and Coatings Technology*, **163** 67-74 (2003).

23. D. G. Cahill, S. K. Watson, and R. O. Pohl, "Lower limit to the thermal conductivity of disordered crystals," *Phys Rev B*, **46** [10] 6131-6140 (1992).

24. T. Hashimoto, J. Kuwahara, T. Yoshidaa, M. Nashimoto, Y. Takahashi, K. Takahashi, Y. Morito, "Thermal conductivity of negative-thermal-expansion oxide, $Zr_{1-x}Y_xW_2O_8$ (x= 0:00; 0.01)- temperature dependence and effect of structural phase transition," *Solid State Communications* **131**, 217-221(2004)

25. J. Wang, H. P. Li, and R. Stevens, "Hafnia and hafnia-toughened ceramics," *J. Mater. Sci.*, **27** [20] 5397-5430 (1992).

26. P. Camagni, G. Samoggia, L. Sangaletti, F. Parmigiani, and N. Zema, "X-ray-photoemission spectroscopy and optical reflectivity of yttrium-stabilized zirconia," *Phys. Rev. B*, **50** [7] 4292

(1994).

27. B. Y. Wang, X. D. Yuan, X. D. Jiang, X. T. Zu, Y. J. Guo, and W. G. Zheng, "The optical properties of SiO$_2$ and ZrO$_2$ films investigated by spectroscopic ellipsometry," *Piezoelectrics & Acoustooptics,* **39** [6] 747-750 (2008).

28. P. J. Martin, R. P. Netterfield, and W. G, "Sainty, Modification of the optical and structural properties of dielectric ZrO$_2$ films by ion-assisted deposition," *J. Appl. Phys.*, **55** [1] 235-241 (1984).

29. Y. Yamamura, N. Nakajima, T. Tsuji, M. Koyano, Y. Iwasa, S. Katayama, K. Saito, and M. Sorai, "Low-temperature heat capacities and Raman spectra of negative thermal expansion compounds ZrW$_2$O$_8$ and HfW$_2$O$_8$," *Phys Rev B*, **66** [1] 014301 (2002).